\documentclass[aps,pra,twocolumn,superscriptaddress]{revtex4}
\usepackage{graphicx,amsmath,amssymb}
\usepackage{bm}
\usepackage{physics}
\usepackage{mathtools}
\usepackage{bm} 
\usepackage{amsthm} 
\usepackage{amsfonts}
\usepackage{hyperref}
\newtheorem{theorem}{Theorem}
\theoremstyle{definition}

\graphicspath{ {/images/} }
\usepackage{xcolor}

\newcommand{\dcq}{\hbox{D}_{\hbox{\tiny QC}}}
\newcommand{\odcq}{\overline{\hbox{D}}_{\hbox{\tiny QC}}}
\begin{document}
\title{Quantum-classical dynamical distance and quantumness of quantum walks}
\author{Valentina Gualtieri}
\author{Claudia Benedetti}\email{claudia.benedetti@unimi.it}
\author{Matteo G. A. Paris}\email{matteo.paris@fisica.unimi.it}
\affiliation{Dipartimento di Fisica "Aldo 
Pontremoli", Universit\`a degli Studi di Milano, I-20133 Milano, Italy}
\date{\today}
\begin{abstract}
We introduce a fidelity-based measure $\dcq (t)$ to quantify the 
differences between the dynamics of classical (CW) and quantum (QW) 
walks over a graph. We provide universal, graph-independent, 
analytic expressions of this quantum-classical dynamical 
distance, showing that at short times 
$\dcq (t)$ is proportional to the coherence of the walker, i.e. a 
genuine quantum feature, whereas for long times it depends only on 
the size of the graph. At intermediate times, $\dcq (t)$ does  
depend on the graph topology through its algebraic connectivity. Our 
results show that the difference in the dynamical behaviour 
of classical and quantum walks is entirely due to the emergence 
of quantum features at short times. In the long time
limit, quantumness and the different nature of the generators 
of the dynamics, e.g. the open system nature of CW and the unitary nature 
of QW, are instead contributing equally. 
\end{abstract}
\maketitle
\section{Introduction}\label{s:intro}
Classical and quantum walks provide  powerful tools to describe
the transport of charge, information or energy in several systems 
of interest for a wide spectrum of disciplines, ranging from 
quantum computing to biological physics \cite{Venegas12,mulken11,
ambainis03,childs09}.
In these contexts, in 
order to understand the very nature of the underlying  dynamics, 
a question often arises on how to compare and assess the different 
behaviors of classical and quantum walks on a given structure. 
Quantum walks are also very useful to build quantum algorithms 
\cite{gut98, ambainis07,childs2004,tamascelli14}, 
and a comparison with the corresponding classical random walks is 
crucial to assess the possible quantum enhancement due to the 
faster spreading of probability distributions. As a consequence, 
the differences between classical and a quantum walk have been 
analyzed quite extensively, with short-and long-time behavior 
studied in both scenarios \cite{childs02,konno05,facc13,reza18,Szigeti19,kop19}. 
Signatures of the nonclassicality of the evolution involve the 
ballistic propagation of the quantum walker, compared to the 
classical diffusive analogue  \cite{mulken}, and their 
measurement-induced disturbance or the presence of non-classical 
correlations, i.e. discord, in bipartite systems \cite{rao11}. 
The effects of classical noise on the gradual loss of quantum features 
has been also investigated \cite{ben1,ben2}
\par
Classical and quantum walkers evolve indeed differently over a 
given graph. In particular, classical random walks are open systems 
where randomness may be ascribed to the interaction with some external 
source of noise,  whereas the evolution of a quantum walker is 
unitary. A crucial question thus arises on whether the different 
behaviour of classical and quantum walks corresponds to the appearance 
of some genuine quantum feature or it is just due to the different 
nature of their dynamics. In order to answer the question, we here 
introduce and discuss a fidelity-based measure, denoted by $\dcq(t)$, 
to quantify the difference between the dynamics of a classical walker 
on a given graph and that of the corresponding quantum walker. We 
discuss some universal properties of our measure, and provide analytic
expressions for short and long times. Our results show that 
at short times the difference is indeed due 
to the appearance of a quantum feature, i.e. coherence, whereas in the 
long times limit quantumness plays only a partial role. In this 
regime, $\dcq$ also contains a term given by 
the distance between the probability distributions 
over the graph and, overall, it depends only on the size of the 
graph. As we will see, the behaviour of $\dcq$  at 
intermediate  times does instead depend on the graph topology 
through its algebraic connectivity.
\par
Continuous-time quantum walks (CT-QW) are usually introduced as the 
quantum generalization of continuous-time Markov chains, also 
called classical random walks (CT-RW). However, while the classical 
random walk is described though the evolution of a {\em probability 
distribution}, governed by a transition matrix (thus being an open system
by construction), the CT-QW dynamics is unitary with the Hamiltonian, 
given by the graph Laplacian, governing the evolution of 
the {\em probability amplitudes} \cite{fahri98}.
Moreover,  for regular lattices (i.e. graphs where each vertex 
has the same number of neighbors) the graph Laplacian is  the discrete 
version of the  continuous-space Laplacian  thus it describes the 
evolution of a free particle on a discretized space
\cite{wong16}. On the contrary, 
for more general and complex graphs, the graph Laplacian cannot be straightforwardly associated to the classical Hamiltonian of a 
free particle. 
\par
The paper is structured as follows: In Section \ref{s:qcd} we 
introduce the notion of QC-distance and prove that the involved 
maximization problem may be solved exactly. In Section \ref{s:uni} 
we discuss the behaviour of the QC-distance at short and long times, 
deriving asymptotic, graph-independent expressions, whereas in Section 
\ref{s:gdp} we instead discuss some of its graph dependent features.
Section \ref{s:role} is devoted to analyze quantitatively the role 
of coherence and classical fidelity in determining the value of the 
QC-distance. Section \ref{s:outro} closes the paper with some 
concluding remarks.
\section{Quantum-classical dynamical distance}\label{s:qcd}
Let us consider a  finite undirected graph $G(V,E)$, where $V$ is the 
set of vertices and $E$ the set of edges. The state of the classical walker at 
a given time is described  by the probability vector 
$\vec{p}(t)=e^{\nu Lt}\vec{p}(0)$, where $\vec{p}(0)$ is the 
initial probability distribution over the vertices, $\nu$
is the transition rate and $L$ is the transfer matrix, also known 
as the Laplacian of the graph \cite{newman}, i.e. a symmetric matrix 
whose rows (or columns) sum to zero. In particular, $L_{jk}=1$ 
(with $j\neq k$) if the nodes $j$ and $k$ are connected by an 
edge and $L_{jk}=0$ if they are not. The diagonal elements of $L$
are given by $L_{jj}=-d_j$, where $d_j$ is the degree of node $j$, i.e.
the number of edges connecting $j$ to other nodes. Given an  initially 
localized probability distribution over the site $j$, and using a quantum mechanical notation, the evolution of a CT-RW may be 
described by the mixed state   
\begin{equation}
\mathcal{E}_{\textsc c}(\rho_j)=\sum_k p_{kj}(t)\ketbra{k}{k},
\label{rhoc}
\end{equation}
where $p_{kj}(t)=\bra{k}e^{\nu Lt}\ket{j}$ is the transition 
probability  from site $j$ to site $k$, $p_{kj}(0)=\delta_{kj}$, 
and the initial localized state is $\rho_j=\ketbra{j}{j}$. 
The orthonormal basis $\{\ket{k}\}_{k=1}^N$ describes localized states of 
the walker on one of the $N$ sites of the graph. The 
completely-positive map $\mathcal{E}_{\textsc c}$ describes the 
dynamics of the CT-RW.
An initially localised quantum walker evolves instead unitarily, 
and the evolved state is given by the pure state 
\begin{align}
\mathcal{E}_{\textsc q}(\rho_j)=\ketbra{\psi_j(t)}{\psi_j(t)},
\quad\ket{\psi_j(t)}=\sum_k \alpha_{kj}(t) \ket{k} 
\label{rhoq}
\end{align}
where the coefficients 
$\alpha_{kj}(t) = \mel{k}{e^{i\nu Lt}}{j}$ 
represent the transition (tunnelling) {\it amplitudes} between 
nodes $j$ and $k$ \cite{fahri98}. 
\par
As it is apparent from Eqs. (\ref{rhoc}) and (\ref{rhoq}) the two 
evolutions lead to completely different final states. First of all, 
the classically evolved state of the CT-QW is always a mixed state, 
while for the CT-QW we have a pure state at all times. In addition, 
quantum evolution admits superpositions of states and interference
effects, which lead to dramatically different evolutions compared to 
the CT-RW. In turn, we remind that in the classical case the 
Laplacian is just the transfer matrix of the Markov chain, whereas 
for CT-QW $L$ is the effective Hamiltonian of the walker, i.e. we have
$H=-\nu L$. Hereafter, and without loss of generality (since it 
corresponds to fixing the time unit), we set the transition rate 
$\nu=1$.
\par
In order to quantify the differences between the classical and the 
quantum dynamics of the walker, and to assess whether they may 
be ascribed to the appearance of genuine quantum features, we introduce 
a fidelity-based measure of dynamical distance (QC-distance) for a 
quantum walker on a graph, and investigate its behavior in time.
The QC-distance $\dcq(t)$ of a quantum walker on a graph $G$ is
defined as
\begin{equation}
\dcq(t) \equiv 1 - \min_{\rho_{\textsc c}} \mathcal{F} \big( \mathcal{E}_{\textsc c} (\rho_{\textsc c}), \mathcal{E}_{\textsc q} (\rho_{\textsc c}) \big),
\label{nonclassicality}
\end{equation}
where $\rho_{\textsc c}$ represents an initial classical state of the
walker, i.e. a diagonal density matrix whose elements
give the initial probability distribution over the graph $G$.
 The quantity  $\mathcal{F} \left( \mathcal{E_C} (\rho_{\textsc c}), \mathcal{E_Q}
 (\rho_{\textsc c}) \right)$ is the quantum  fidelity \cite{Jozsa94,rag01,gil05,chen18} 
between the two states obtained evolving $\rho_{\textsc c}$ using the quantum 
and the classical map, respectively, i.e.
$ \mathcal{F}(\rho_1,\rho_2)=\left[\Tr\sqrt{\sqrt{\rho_1}\rho_2\sqrt{\rho_1}}\right]^2.$
Notice that in the definition \eqref{nonclassicality}
we take the minimum of the fidelity over all initial classical states. 
This is to capture the intuition that the QC-distance should be large 
if at least one classical states is evolving very differently under 
the two dynamical maps. According to its definition, 
the QC-distance $\dcq(t)$ is a positive quantity bounded 
between 0 and 1.
\par
Let us now prove that 
for any graph the initial state that gives 
the minimum in Eq. \eqref{nonclassicality} is a localized state, i.e.
a state of the form $\rho_j=\op{j}$.
\theoremstyle{theorem}
\begin{theorem}
The initial classical state attaining the minimum in Eq. 
\eqref{nonclassicality} is a localized state $\rho_j=\ketbra{j}{j}$.
\label{teo1}
\end{theorem}
\emph{Proof:}
Let us consider a generic classical state $\rho_{\textsc c} = \sum_k z_k \rho_k$, with $\rho_k = \op{k}{k}$. The coefficients $\{z_k\}$ give the 
initial probability distribution of the walker over the graph sites, satisfying the normalization condition $\sum_k z_k =1$.
In order to evaluate the QC-distance of the walker, 
we need  to find the state $\rho_c$  that minimizes the fidelity 
between the evolved CT-RW and the CT-QW, described respectively by the quantum maps $\mathcal{E}_{\textsc c} (\rho_{\textsc c})$ and $\mathcal{E}_{\textsc q} (\rho_{\textsc c})$.
The strong concavity property \cite{uhlmann00} applied to the 
square root of the fidelity gives:
\begin{align}
\sqrt{\mathcal{F} ( {\mathcal{E_\textsc{c}} ( \rho_\textsc{c}), \mathcal{E_\textsc{q}} (\rho_\textsc{c}) })}
& \geq  \sum_k \! z_k \! \sqrt{\!\mathcal{F}( { \mathcal{E}_\textsc{c}(\rho_k), \mathcal{E}_\textsc{q}(\rho_k)})}
\end{align}
where we omitted the explicit dependency on time.
For future convenience let us also introduce the shorthand 
$\mathcal{F}_k=\mathcal{F}(\mathcal{E}_{\textsc c}(\rho_k), 
\mathcal{E}_{\textsc q}(\rho_k))$ for the fidelity between 
the classical and the quantum evolved state of a walker
initially localized on site $k$.
For regular graphs, i.e. graphs where each vertex has the same number of neighbors,
all nodes are equivalent and  the fidelity does not depend on the initial site $k$,
 hence $F_k = F_0$. Therefore, thanks to the monotonicity of the square root 
 and to the normalization condition, we have: i)
 ${\mathcal{F} \left( \mathcal{E_{\textsc c}} (\rho_{\textsc c}), \mathcal{E_{\textsc q}} (\rho_{\textsc c}) \right)} \geq {F_0}$, and ii) the minimum is obtained for an initially localized state. For non-regular graphs, we have the same conclusion since $\sum_k z_k \sqrt{F_k}$ is a convex combination of limited functions, and thus its 
 minimum is given by
\begin{align}
\min_k \sum_k z_k \sqrt{\mathcal{F}_k} = \min_k \sqrt{\mathcal{F}_k}\,,
\end{align}
i.e. it is achieved by an initially localized state.
\hfill $ \blacksquare$
\begin{figure*}[ht]
\includegraphics[width=0.31\textwidth]{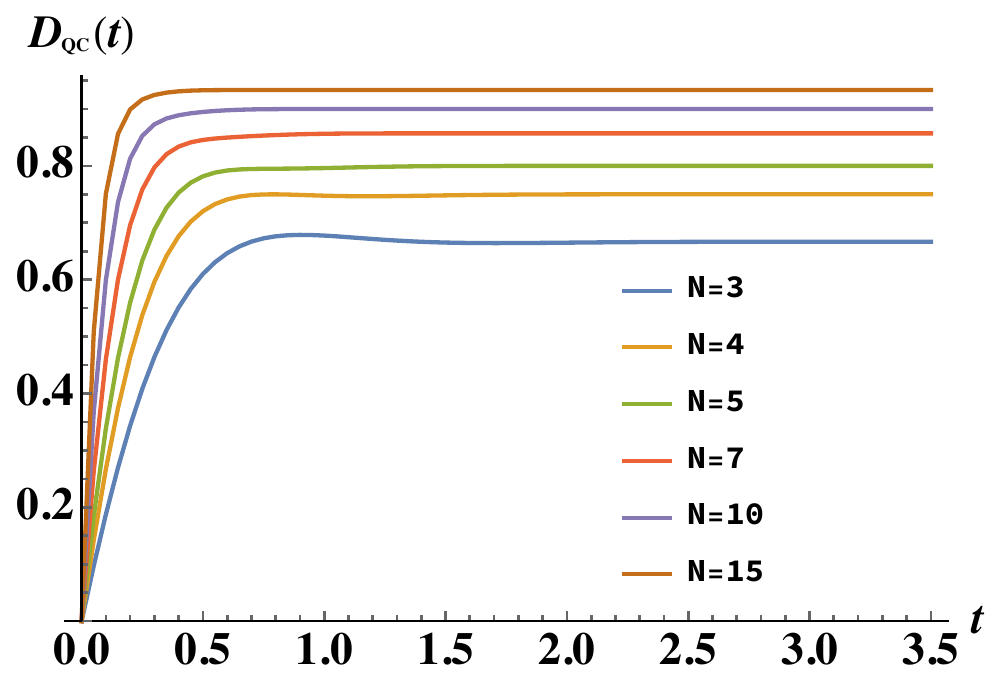}
\includegraphics[width=0.31\textwidth]{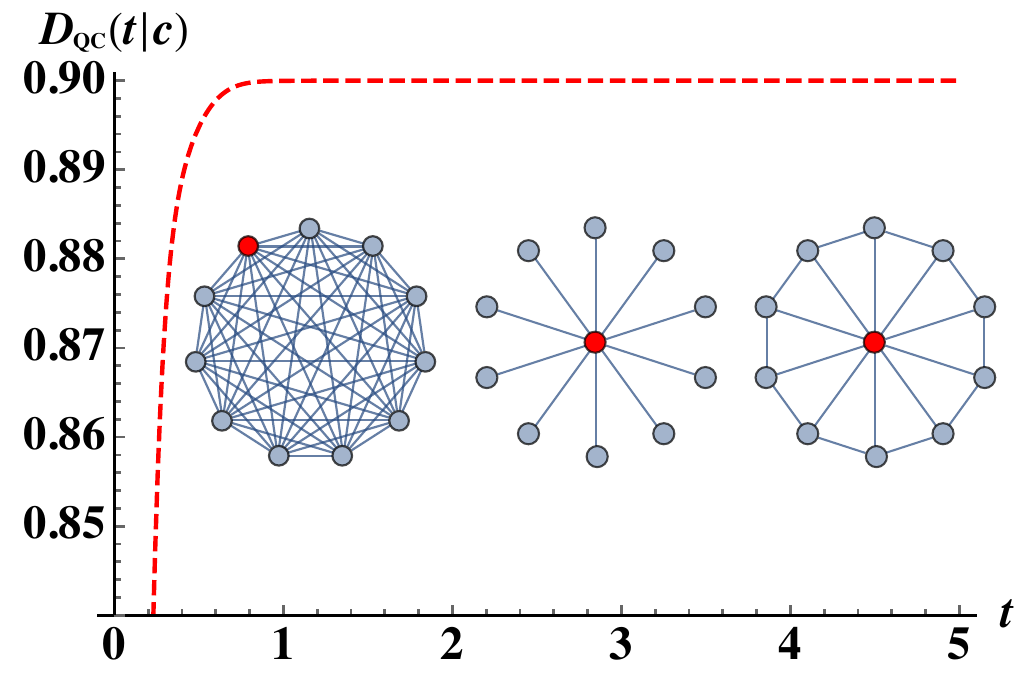}
\includegraphics[width=0.31\textwidth]{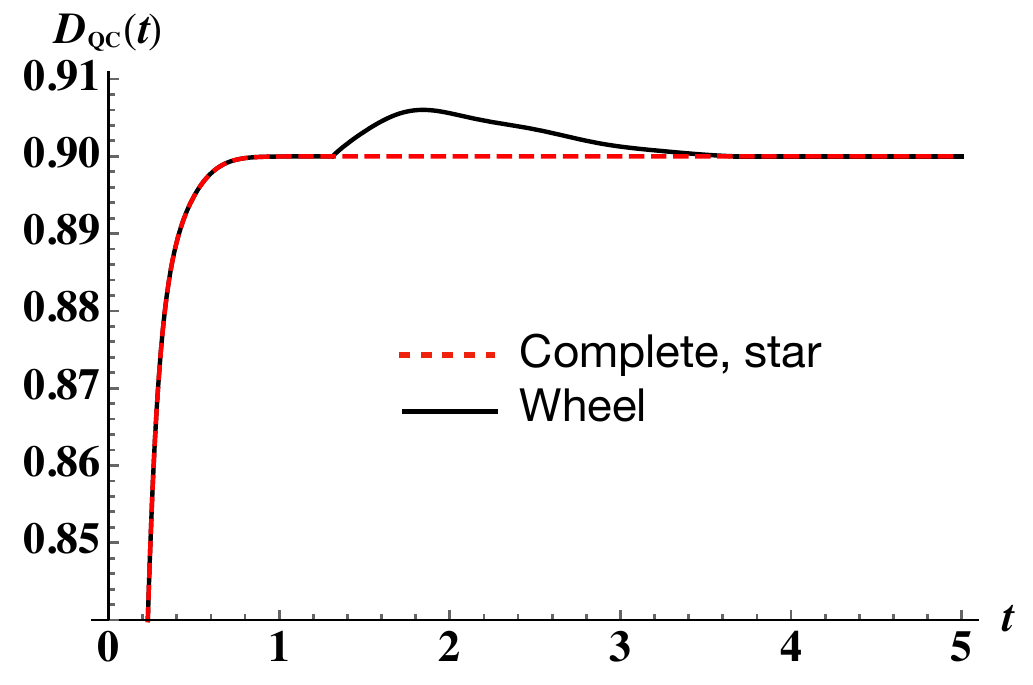}
\caption{Left panel: QC-distance for complete graphs of different sizes $N$.  
Central panel: the quantity $\dcq(t|c)$ for the complete, star and wheel graphs, 
starting from the initial central site $\ket{c}$, depicted in red in the 
insets. Right panel: QC-distance for the graphs of the central panel.}
\label{nccomp}
\end{figure*}
\section{Universal properties of the QC-distance} \label{s:uni}
As already mentioned above, the QC-distance $\dcq(t)$ is a positive
 quantity bounded between 0 and 1. Since we know from theorem \ref{teo1} that 
the optimal initial state achieving the maximum in \eqref{nonclassicality} 
is a localized state $\rho_j$, let us analyze the temporal behavior and 
properties of the fidelity 
\begin{equation}
\mathcal{F}_j(t)=\sum_k p_{kj}(t) |\alpha_{kj}(t) |^2
\end{equation}
for a walker initially localized on the node $j$.
This expression allows us to explore the behavior of the 
conditional distances $\dcq(t|j)=1-\mathcal{F}_j(t)$ in 
different regimes. In particular, in the short-time limit $t\ll1$, we 
find that $\dcq(t|j)$ depends only on the degree of the 
corresponding node, i.e. $d_j=-\bra{j}L\ket{j}$, as follows
\begin{equation}
 \dcq(t \ll 1 |j)=d_j t + O(t^2)\,.
 \label{shortt}
 \end{equation}
 This result is obtained by expanding the transition 
 probabilities $p_{kj}(t)$ and the tunneling amplitudes $\alpha_{kj}(t)$ 
 up to first order in time
 \begin{align}
 p_{kj}(t) & =  \delta_{kj} + t \mel{k}{L}{j} + O(t^2) \\
 \alpha_{kj}(t)  & =  \delta_{kj} +it \mel{k}{L}{j} + O(t^2)\,,
 \end{align}
 and then substituting these expression in $\mathcal{F}_j(t)$, with
 the reminder that the off-diagonal elements of 
 $L$ are positive, while the diagonal ones are negative.
 The meaning of Eq. \eqref{shortt} is that the more connected  
 the initial node is, the larger is the difference betweeen the 
 dynamics of a QW and a RW on the given graph. The QC-distance 
 $ \dcq(t) =\max_j\dcq(t|j)$ for a given graph is thus determined 
 by the vertex with maximum degree. 
 \par
Concerning the behaviour for large times, we notice that for a classical 
walker the distribution over the nodes tends to the flat distribution, 
i.e. for $t \gg 1$ we have $\mathcal{E}_{\textsc c}(\rho_\textsc{c}) = 
\mathbb{I}/N$, $\forall \rho_\textsc{c}$. In turn, we have 
$\mathcal{F}_j=1/N$, $\forall j$ and therefore
we can rewrite the QC-distance in the long-time regime as
\begin{equation}
\dcq(t\gg1) \simeq 1-\frac{1}{N}\,,
\label{longt}
\end{equation}
independently on the topology of the graph.
\par
The physical interpretation of the above results is rather 
clear: at short times what really matters is the connecttivity 
of the initial node. This is a local phenomenon and does not depend 
on the dimension of the graph. As time passes, classical and quantum 
walkers evolve, and explore  the whole graph until the CT-RW achieves 
the stationary uniform distribution over the graph, while the CT-QW 
periodically evolves both in populations and coherences. This leads 
to a stationary value for the QC-distance, depending only on the size 
of the graph, which is a global property.  This is illustrated in 
the left panel of Fig. \ref{nccomp}, where we display, as an example, 
the  behavior of the QC-distance as a function of time for complete 
graphs of different sizes. The initial slope of the curves at short 
times is  the vertex degree, while at long times the stationary value 
$1-1/N$ is reached.
\par
The intermediate-time behavior of $\dcq(t)$ is  related to the topology 
of the graph, with the main contribution coming from its  algebraic 
connectivity. 
In order to see this, we notice that the squared amplitudes
$|\alpha_{kj}(t) |^2$ are bounded (and oscillating) functions, 
whereas the classical transition probabilities may be written as 
\begin{align}
p_{kj}(t) & = \sum_{s=0}^N e^{-|\lambda_s| t} \langle k |\lambda_s\rangle\langle\lambda_s|j\rangle \notag \\ & 
= \delta_{kj}+ \sum_{s=1}^N e^{-|\lambda_s| t} \langle k |\lambda_s\rangle\langle\lambda_s|j\rangle\,,
\end{align}
where we have introduced the eigenvalues and eigenvectors of the Laplacian
$L= \sum_s \lambda_s |\lambda_s\rangle\langle\lambda_s| $ and already took
into account that the smallest (in modulus) eigenvalue 
of a Laplacian is always zero. The dominant term in $p_{kj}(t)$ and, in 
turn, in the fidelity, is thus the one containing $|\lambda_1|$, which 
is usually referred to as the Fiedler value or Fiedler eigenvalue of
the Laplacian, providing an overall algebraic 
quantification of the connectivity of the graph \cite{fiedler}.
\section{Graph-dependent properties of the QC-distance} \label{s:gdp}
The definition of the QC-distance involves a maximization over the 
initial state of the walker. There may be, however, situations where 
the $\dcq(t|j)$ themselves may be of interest, e.g. when there exists 
a privileged node to start with, and we want to assess the effect
of different topologies. This kind of situation is illustrated in the 
central and right panels of Fig. \ref{nccomp}, where we  compare the 
behavior of $\dcq(t|c)$ for the complete, star and wheel graphs,
$\ket{c}$ being the central node (see the red points in the inset). 
As shown in the central panel, $\dcq(t|c)$ is the same for 
all graphs, since   they all have a central node $\ket{c}$ with degree 
 $N-1$  and there is at least a localized 
 preparation on all graphs leading to the same dynamics.   
On the other hand, if we look at the QC-distance 
$\dcq(t)$, we see 
that for the wheel graph it departs from the others curves
in a certain time interval. In fact, it increases linearly at 
short times according to Eq.  \eqref{shortt}, whereas, as time grows, 
the proportionality is lost and the topology of the graph plays 
a role in the behavior of the NC. 
This is physically consistent, since $\dcq(t)$ distance aims to 
 quantify a property of the graph itself rather than the properties
 of specific preparations.
\par
\begin{figure}[h!]
\centering
\includegraphics[width=0.9\columnwidth]{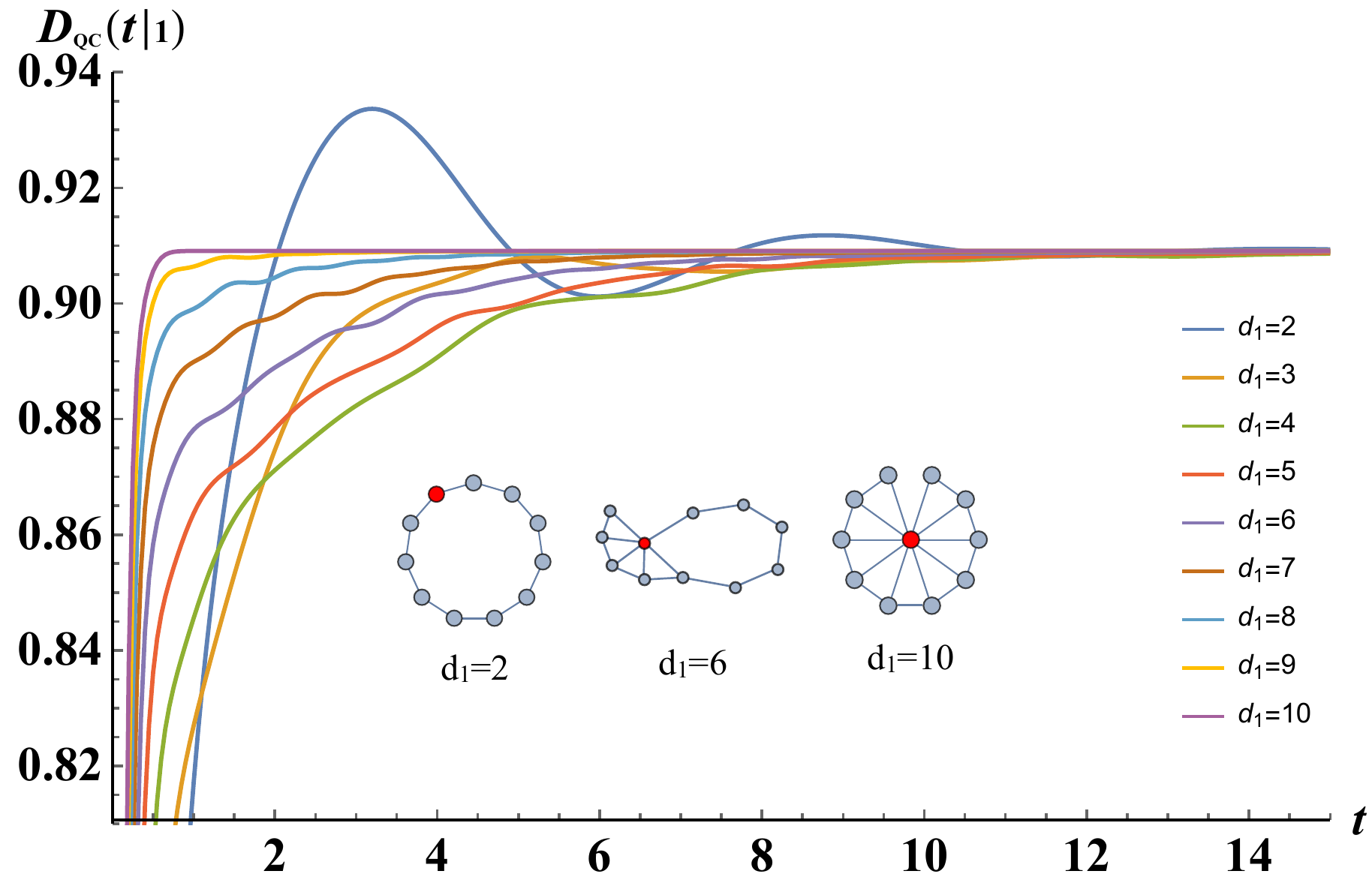}
\caption{The distance $\dcq(t|1)$  for different graphs 
having the same size $N=11$ and different degree of node $1$.
The blue oscillating line denotes $\dcq(t|1)$ for the ring 
graph. The insets show some of the considered graphs with 
degrees $d_1=2,6,10$ respectively. The plot shows that for
short times the ring graph has the lowest value of $\dcq(t|1)$,
whereas at later times its $\dcq(t|1)$ is larger than those of  
the other graphs, i.e. it departs more from the classical
dynamics.
}
\label{degr}
\end{figure}

Let us illustrate this behaviour with a different example, i.e.
we consider different graphs with a fixed size, say $N=11$, and
different connectivity. In particular, we start by consider a 
a ring graph, where all the nodes have degree equal to two and
then select one node, e.g. $\ket{1}$ and take random connected 
graphs with increasing number of links, i.e. we increase 
the node degree $d_1$. 
The behaviour of  $\dcq(t|1)$ is shown in 
Fig. \ref{degr}. At short times, the ring graph has the lowest 
value of $\dcq(t|1)$, but then it shows a maximum value in time, 
which is higher compared to the other graphs. In other words, 
the evolution of a quantum walker on a ring graph is initially 
closer to its classical counterpart compared to other graphs 
with larger $d_1$, but then, for larger times, it becomes 
more nonclassical, i.e. it departs more from the classical
dynamics compared to the other considered graphs. The insets 
show some of the considered graphs with degrees $d_1=2,6,10$, 
respectively.
\par
Depending on the application at hand, one may be also interested in 
assessing the average dynamics over a graph. To this aim, let us also
briefly discuss another notion of QC-distance, taking into account 
the role of different initial positions. This is the average of 
$\dcq(t|j)$ over the localized states, i.e. 
\begin{align}
\odcq(t)=\frac{1}{N}\sum_{k=1}^N \dcq(t|j),
\end{align}
which may be naturally referred to as the {\em average QC-distance}. 
For regular graphs, it coincides with $\dcq(t)$, whereas for 
non-regular graphs it accounts for the fact that a walker initially 
localized on different nodes may evolve very differently.  
The behaviour of $\odcq(t)$ 
may be easily  recovered from the previous analysis. We have 
${\odcq}(t\gg 1) \simeq \overline{d}\, t$ for short times, 
where $\overline{d}$ is the average degree of the graph 
and $ {\odcq}(t)(t\gg1)\simeq 1-\frac{1}{N}$ for long times.
\par
\begin{figure*}[ht]
\centering
\includegraphics[width=0.4\textwidth]{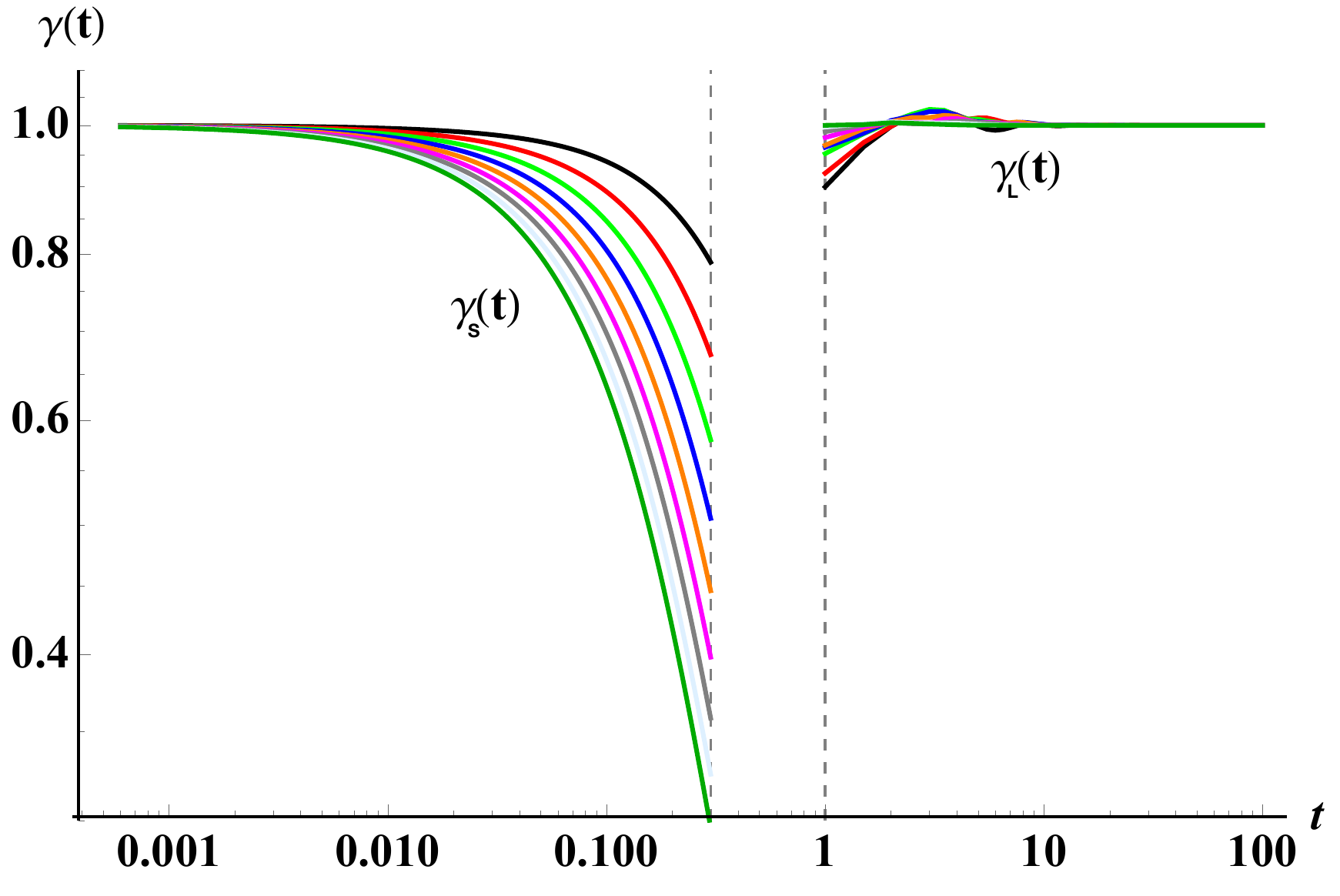}$\quad$
\includegraphics[width=0.4\textwidth]{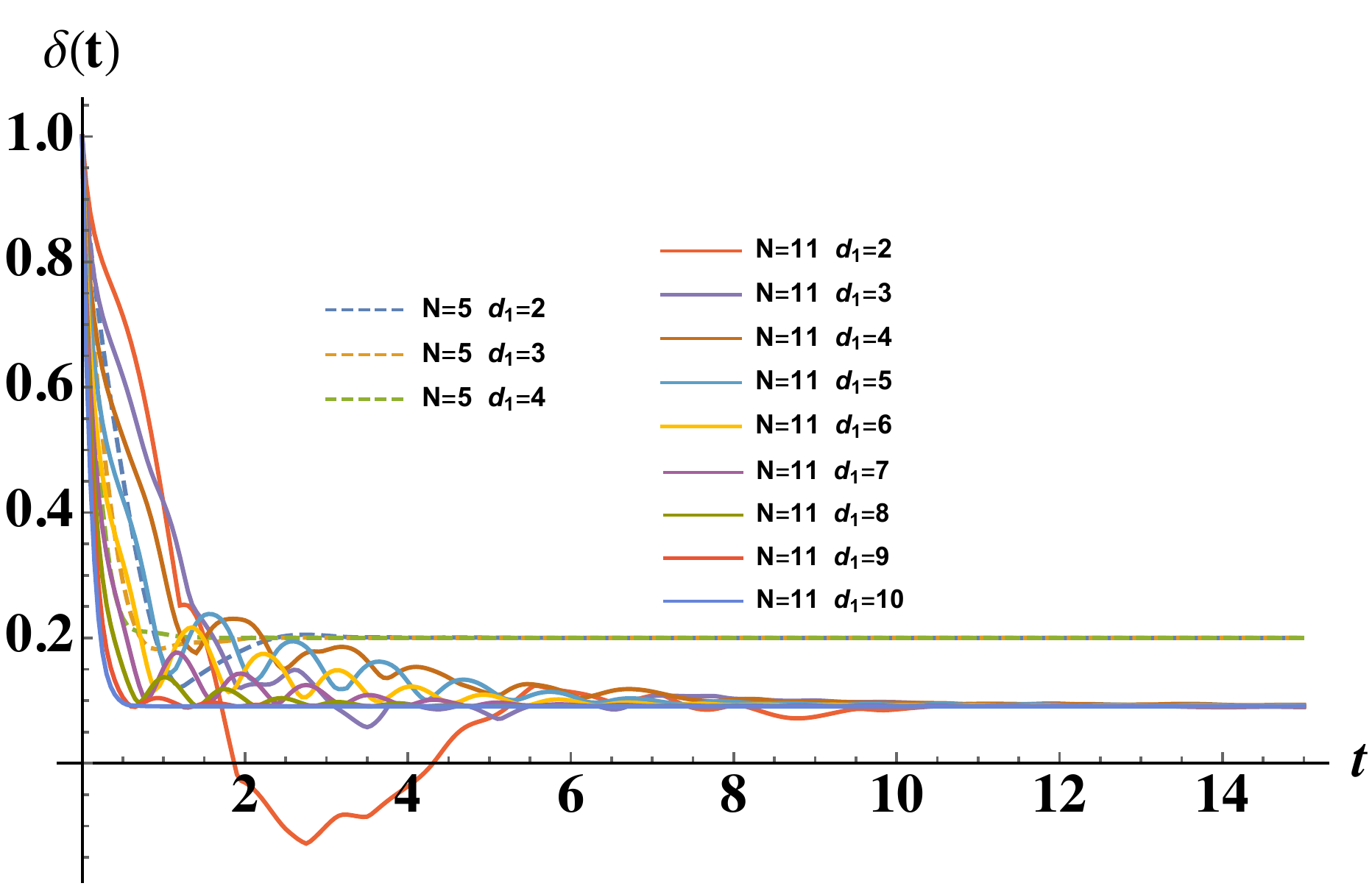}
\caption{Left panel: the ratios $\gamma_{\text {\tiny S}}(t)$ and 
$\gamma_{\text {\tiny L}}(t)$ between the exact QC-distance 
and its asymptotic expressions derived from \eqref{final_cfr} for short and 
long times for several graphs of size $N=11$. The plot illustrates the fact 
that the range of validity of the short time expression 
$ \dcq^{\text{\tiny S}}(t|j)$ depends quite 
strongly on the kind of graph, whereas the convergence to the 
asymptotic value $\dcq^{\text{\tiny L}}(t)=1-1/N$ is almost 
independent on the  graph, and it is achieved quite rapidly.
Right panel: the difference 
$\delta(t)$ 
between the classical distance and the renormalized coherence for
 the several graphs of size $N=11$ and $N=5$.
 Here the convergence time increases with the size of 
the graph, still being independent on its topology.
 }
\label{gammas}
\end{figure*}
\section{The role of coherence and classical fidelity} \label{s:role}
The QC-distance quantifies how much the evolution of a
quantum walker on a graph differs from its CT-RW counterpart. 
A question arises on whether this difference is due to the 
appearance of  genuine quantum features, or it is just due to 
differences in the two maps $\mathcal{E}_{\textsc c}$
and $\mathcal{E}_{\textsc q}$.
As we will see  the answer is not trivial and time-dependent. 
Let us briefly recall the notion of coherence of a quantum 
state, a genuine quantum property with no classical analogue. 
Coherence  may be properly quantified by the sum of 
the off-diagonal elements of the density matrix, i.e. \cite{plenio14}
$C(t) \equiv \sum_{k \neq j} \abs{\rho_{kj}(t)}$. For the dynamics
of a quantum walker the natural basis to consider is that of localized 
states. The coherence at time $t$ is thus given by 
\begin{align}
C_j (t) = \left( \sum_k \abs{\alpha_{kj}(t)} \right)^2 - 1,
\end{align}
where the index $j$ refers to the localized initial state 
of the quantum walker. By construction, any classical state 
of the form \eqref{rhoc} has zero coherence, i.e. it is incoherent. 
By expanding this expression for short times, up to first order, 
and comparing it with the expression of nonclassicality in Eq.  
\eqref{shortt} we find $$ \dcq(t\ll1|j)=\frac12 C_j(t)\,.$$ 
It follows that the initial behavior of  the QC-distance 
at short times is governed solely by the amount of  coherence 
created by the dynamics. In other words, the difference in the dynamics
may be fully attributed to the appearance of genuine quantum features. 
On the other hand, this is no longer true at later times, where 
a substantial contribution to $\dcq(t)$ is due to differences in the distribution
over sites. In order to prove this statement, let us introduce the
classical fidelity between the  probability distributions over the 
sites of CT-RW and CT-QW, i.e. 
\begin{equation}
 G_j(t)=\sum_k\sqrt{p_{kj}(t)|\alpha_{kj}(t)|^2}\,.
 \end{equation}
For large times $p_{kj}(t)\simeq 1/N$, and thus we have $$\sqrt{N} G_j(t) {\simeq}
\sum_k |\alpha_{kj}(t)|\,$$ and, in turn, 
$$N\,G_j^2(t)-C_j(t){\simeq}1\,.$$
Since for large times $\dcq(t|j) \simeq 1 -1/N$, we may summarize 
the above results as follows
\begin{equation}
 \dcq(t|j)=\begin{cases}
 \dcq^{\text{\tiny S}}(t|j)\equiv\frac12\, {C_j(t)} & t\ll 1\\
 & \label{final_cfr} \\
  \dcq^{\text{\tiny L}}(t|j)\equiv1-G_j^2(t)+\frac1N\,{C_j(t)} & t\gg1
 \end{cases}\,,
 \end{equation}
from which, after maximizing over nodes, we obtain the asymptotic expression 
of $\dcq(t)$ in terms of coherence and classical fidelity.
Eq. (\ref{final_cfr}) shows that that for 
short times a nonzero QC-distance may be ascribed to the appearance 
of coherence, whereas for long times quantum features accounts only 
partially for the difference between the two dynamics. In this regime, 
QC-distance is the sum of the normalised coherence and the 
distance between the probability distributions over the nodes of
the graph. We also remark that $\dcq(t)$ no longer depends on the topology 
of the consider graph, but rather only on its size. In order to assess
the generality of this statement, and the range of validity  of
Eq. \eqref{final_cfr}, we have considered different classes of graphs
and evaluated the ratios 
$$\gamma_{\text {\tiny K}}(t)=\dcq(t)/\dcq^{\text{\tiny K}}(t)\, 
\qquad K=S,L\,,$$ between the exact QC-distance 
(calculated numerically) and its limiting expressions derived from
Eq. (\ref{final_cfr}) for short and long times. In the left panel 
of Fig. \ref{gammas} we report the two values of $\gamma$ for a set of 
random graphs of size $N=11$. As it is apparent from the plot, the range 
of validity of the short time expression $ \dcq^{\text{\tiny S}}(t|j)$ 
depends quite 
strongly on the kind of graph, whereas the convergence to the 
asymptotic value $\dcq^{\text{\tiny L}}(t)=1-1/N$ is almost 
independent on the  graph, and it is achieved quite rapidly. The
same rapid convergence to the value $\delta(t)= 1/N$ 
may be seen for the difference $\delta(t)=G^2(t)-C(t)/N$ 
between the square  of the  classical fidelity and the size-normalized 
coherence (see the right panel of Fig. \ref{gammas}). 
Here the convergence time increases with the size of 
the graph, still being independent on its topology.
\section{Discussion and conclusions} \label{s:outro}
We have introduced a fidelity-based measure, termed QC-distance 
$\dcq(t)$, to properly compare the dynamical behaviour of 
classical and quantum walks over a graph, also discussing the role 
of size and topology of the graph. Our results show 
that at  short times, the QC-distance of quantum walks is 
proportional to the local connectivity, and in turn to 
coherence, i.e. to the appearance of a genuine quantum 
feature. On the other hand, in the long time limit, 
quantumness plays only 
a partial role, since the QC-distance is the sum of
a size-normalised measure of coherence and the 
classical distance between the probability 
distributions over the graph. The graph topology is not 
relevant in those two limiting regimes, whereas it plays 
a role in determining the QC-distance at intermediate 
times. Notice that the two terms in $\dcq^L(t)$ are 
approximately of the same magnitude, i.e. coherence 
and classical distance contribute almost equally to the 
QC-distance.
\par
From the physical point of view, the behavior of $\dcq(t)$ 
tells us that the difference between CT-RW and CT-QW may be 
initially ascribed to the ability of a quantum walker to 
{\em tunnel} between sites, whereas for longer times 
coherence cannot fully account for the difference in 
the dynamics. In this regime, QC-distance is also due 
to the periodic nature of CT-QW dynamics, compared to the 
diffusive one of CT-RW, which leads to an equilibrium state. 
In other words, the differences in the long times dynamics 
should be equally ascribed to the appearance of quantum 
features, as well as to the different nature (open vs 
closed system) of the two dynamical models.
\par
We put forward our measure as a tool in assessing 
the role of quantum features in the dynamics of quantum 
complex networks and in the design of quantum protocols
over graphs. We also believe that it paves the way 
to define the nature and the amount of quantumness
in many particle quantum walks.
\section*{Acknowledgements}
MGAP is member of INdAM-GNFM. We thank Sahar Alipour, 
Gabriele Bressanini, and Ali Rezhakani for useful 
discussions.

\end{document}